\documentclass[11pt]{article}
\usepackage{moriond,epsfig}
\usepackage{epsfig}

\def\Journal#1#2#3#4{{#1} {\bf #2}, #3 (#4)}

\def\PRL{\em Phys. Rev. Lett.}

\def\EPJC{{\em Eur. Phys. Jour.} C}
\def\JHEP{{\em JHEP}}

\def\be{\begin{equation}}
\def\ee{\end{equation}}
\def\bea{\begin{eqnarray}}
\def\eea{\end{eqnarray}}

\newcommand{\as}                {\ensuremath{\alpha_\mathrm{S}}}
\newcommand{\asmz}              {\ensuremath{\alpha_\mathrm{S}(M_{\mathrm{Z^0}})}}
\newcommand{\epem}              {\ensuremath{\mathrm{e^+e^-}}}

\newcommand{\resultnnlo} {\ensuremath{\asmz=0.1201\pm0.0008(\mathrm{stat.})\pm0.0013(\mathrm{exp.})\pm0.0010(\mathrm{had.})\pm0.0024(\mathrm{theo.})}}
\newcommand{\resultnlla} {\ensuremath{\asmz=0.1189\pm0.0008(\mathrm{stat.})\pm0.0016(\mathrm{exp.})\pm0.0010(\mathrm{had.})\pm0.0036(\mathrm{theo.})}}

\begin{document}
\vspace*{4cm}
\title{Determination of \boldmath{\as} using hadronic event shape
  distributions \\ of data taken with the OPAL detector}

\author{ J. Schieck \\  for the OPAL Collaboration}

\address{Ludwig-Maximilians-Universit\"at
  M\"unchen , Am Coulombwall 1, 85748 Garching, Germany  and\\
Excellence Cluster Universe, Boltzmannstrasse 2, 85748 Garching, Germany}

\maketitle\abstracts{
The measurement of the strong coupling \as\ using hadronic event shape
distributions measured with the OPAL detector at center-of-mass
energies between 91 and 209 GeV is summarized. For this measurement
hadronic event shape 
distributions are compared to theoretical predictions based on
next-to-next-to-leading-calculations (NNLO) and NNLO combined with
resummed next-to-leading-logarithm calculations (NLLA). The combined 
result using NNLO calculations is \resultnnlo\ and the result using
NLLO and NLLA calculations is \resultnlla, with both measurements
being in agreement with the world average.
}
\section{Introduction}
The annihilation of electron-positron pairs to  hadronic final
states offers a clean environment to study the theory of the strong
interaction, Quantum Chromo Dynamics (QCD). In particular  
hadronic event shape distributions can be used to measure
the strong coupling \as. \par
During data-taking of the four LEP-experiments 
only next-to-leading order calculations (NLO)
combined with resummed NLLA predictions were available, leading to 
a theoretical uncertainty in the \as\ measurement dominating the
overall uncertainty by far. Only recently new theoretical
calculations become available~\cite{TheoGehrmann,TheoWeinzierl}, which
 take additional  $\as^{3}$  loop corrections into account, so-called NNLO
 calculations. For this analysis these NNLO predictions are used to
 determine \as\ using data taken with the OPAL detector at LEP. Also
 the matched NNLO+NLLA calculations are used. This note gives an overview
 of the analysis performed by the OPAL collaboration. The complete
 description with all details can be found at~\cite{OPALPaper}.
\subsection{Data Sample, Monte Carlo Sample and Event Selection}
\label{MonteCarlo}
We use data taken with the OPAL detector at LEP at center-of-mass
energies between 91 and 209 GeV. Data was taken at 13
different energy points with different event statistics. The largest
event statistics of several hundred thousand events is available at 91
GeV,  due to the large cross-section at the
$Z^{0}$-Resonance. At higher energies only few hundred and at most 
three thousand events are selected. For clarification we group the
result in four different energy intervals with  mean energies of 91, 133, 177 and
197~GeV. \par
For correction of acceptance and resolution effects as
well as for the simulation of the transition from partons to hadrons
a large sample of Monte Carlo events based on the Pythia, Herwig and
Aridane  is generated.
Pythia is used as the default choice and the other event generators are used to estimate the
systematic
uncertainty. The validity of the Monte Carlo models is tested with a 
comparison between the theoretical NNLO calculations and the Monte Carlo
predictions at parton level. The difference between the
NNLO prediction and the Monte Carlo model is well covered by
using the different Monte Carlo as systematic uncertainty. \par
For the analysis well measured hadronic events are selected. For data
taken above the $Z^{0}$-resonance events with large initial state
radiation are removed. Above the $W$-pair threshold the expected
contribution from this four-fermion processes are removed.
\section{Results}
\subsection{Fit procedure}
To measure the strong coupling \as\ event shape observables are
built from selected hadronic events together with using the theoretical
predictions. These observables are constructed in a way that they 
show a large sensitivity to the strong coupling \as. The theoretical prediction is
then fitted to the data distribution with \as\ being the only free
parameter. The following event shape observables are used:
Thrust, heavy jet mass, the total and the wide jet
broadening, the C-parameter and the two-three transition parameter
using the Durham jet algorithm. 
The fit range is determined by requiring the corrections to
be small as well as the theoretical predictions to be stable within
the fit range. 
To compare this analysis with the previous analysis the data is also
fitted to next-to-leading (NLO) and NLO combined with resummed NLLA
calculations. The result from the previous analysis can be reproduced.\par
To asses the systematic uncertainty the fit is repeated in slightly
different ways. Besides the uncertainty due to the correction for
hadronization effects, as described in~\ref{MonteCarlo}, uncertainties
due to the experimental technique and uncertainties due to the
incomplete power series of the theoretical prediction are
evaluated. The overall uncertainty is completed by the statistical
uncertainty originating from the finite statistics used in the
analysis. The main motivation for the re-analysis of the data is the
availability of improved theoretical calculations. In the past this
uncertainty related to the theoretical prediction dominated the overall
uncertainty. However, even with the new improved theoretical
predictions the overall uncertainty is still dominated by the theoretical
uncertainty. The statistical uncertainty, the experimental uncertainty and the
hadronization uncertainty are similar, while the theoretical
uncertainty is at least twice as large.
\subsection{Combination of results}
A single value of the strong coupling \as\ is measured for each event
shape observable and for each energy interval separately.
In order to obtain a single value for each event shape observable or
at each energy interval the values of \as\ are combined. 
The correlation between the different event
shape observables and the different energy intervals are taken
into account. \par
\begin{figure}[ht]
\begin{center}
 \epsfig{file=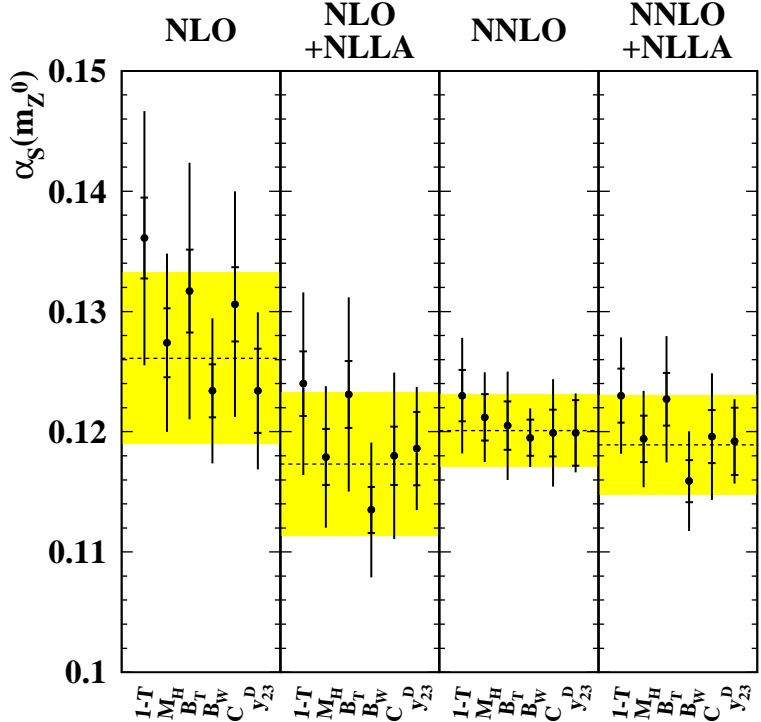, scale=0.5}
  \caption{The measured value of \as\ for the different event shape
    observables combined over the complete center-of-mass energy
    range. The inner uncertainty bar corresponds to the statistical
    uncertainty the outer one to the total uncertainty. The yellow band
   represents the combined \as-value an its uncertainty for all 
    event shape observables using
    NLO, NLO+NLLA, NNLO and NNLO+NLLA calculations.}
\label{EVScatter}
\end{center}
\end{figure}

The combined result for each event shape observable is shown in
Fig.~\ref{EVScatter}. It can be clearly seen that the scatter of
the \as-values obtained with different event shape observables
using NNLO predictions is smaller compared to the measurement using NLO
predictions only. In addition it can be observed, that the overall
uncertainty is reduced with including higher order predictions in the
measurement. The increase of the uncertainty between NNLO and matched
NNLO+NLLA calculations can be explained by the fact that the NNLO
renormalization scale variation is compensated in two loops, while 
the NLLA renormalization scale variation compensation is only 
in one loop. \par
The final result combining all event shape observables at all energy
intervals is \resultnnlo\ using NNLO calculations only and \resultnlla\
using combined NNLO+NLLA predictions.
\subsection{Renormalization scale dependence and running of  \as}
The dependence of the result on the choice of the renormalization
scale is studied. The fixed-order predictions return for the 
fit the smallest $\chi^{2}$-values  at a very small renormalization
scales, while using the matched NNLO+NLLA predictions smaller 
 $\chi^{2}$-values for larger renormalization scales are returned. \par
Together with the re-analyzed JADE result~\cite{JADE} this analysis confirms the
running of the strong coupling \as\ with center-of-mass energy, as
predicted by QCD.
\section{Summary}
\begin{figure}[ht]
\begin{center}
  \epsfig{file=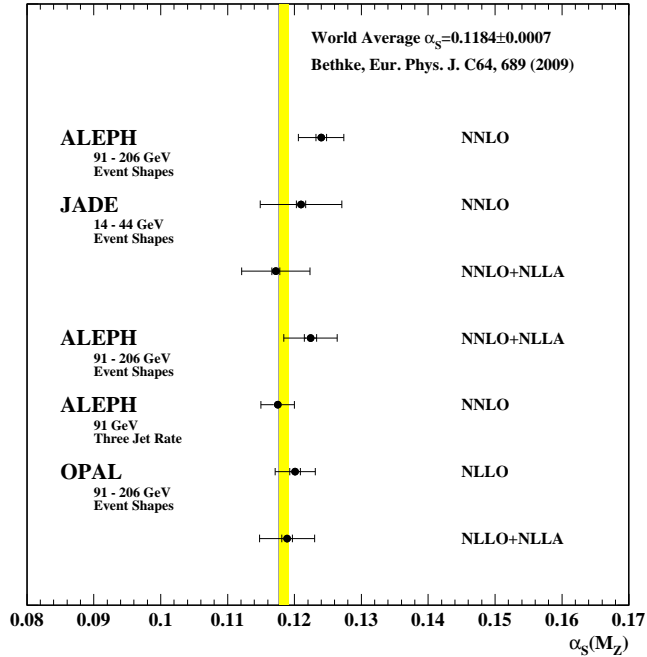, scale=0.5}
  \caption{The combined value of the strong coupling \as\ compared
    to similar measurements also using NLLO and matched NLLO+NLLA
    calculations. The yellow band indicates the world average of the
    strong coupling \as.}
\label{ResultComp}
\end{center}
\end{figure}

The availability of  NNLO predictions for the
annihilation of an electron-positron-pair into a pair of quarks lead
to a re-analysis of data taken with the OPAL detector.
 The combined value obtained for the strong coupling using
NNLO+NLLA calculations results
to \resultnlla, with the overall uncertainty being dominated by
missing higher order terms in the theoretical prediction.  The result
is consistent with the world average~\cite{Bethke}.
The result obtained can be compared to similar analyses 
using NNLO- and NNLO-calculations
matched with NLLA ~\cite{JADE}~\cite{ALEPHNNLO}~\cite{ALEPHNLLA}~\cite{ALEPHJET}.
A summary of these results is shown in Fig.~\ref{ResultComp}. As seen
in this analysis the results obtained using  NNLO+NLLA predictions lead to 
smaller \as-value compared to a pure NNLO fit.  The smallest overall
uncertainty is obtained with a fit to the three-jet
rate.
Several ways to measure the strong coupling \as\
using \epem-data do exist. The value obtained using $\tau$-decays or applying
a fit to electroweak precision observables result in a smaller overall
uncertainty compared to this measurement~\cite{Bethke}.
Besides a precise determination of a fundamental parameter of the
Standard Model this measurement can be seen as a consistency check of
the theory of strong interactions, Quantum Chromo Dynamics.
\section*{Acknowledgments}
This research was supported by the DFG cluster of excellence 'Origin
and Structure of the Universe'.

\end{document}